\documentclass[trackchanges,twocolumn]{aastex701}

\usepackage{graphicx}
\usepackage{longtable}
\usepackage{array}
\usepackage[version=4]{mhchem}

\begin{document}

\title{Mapping Stellar Heterogeneities with the Nautilus Space Observatory}

\author[orcid=0000-0002-9464-8101,sname='A.~D.~Fenstein']{Adina~D.~Feinstein}
\affiliation{Department of Physics and Astronomy, Michigan State University, East Lansing, MI 48824 USA}
\email[show]{adina@msu.edu} 


\author[orcid=0000-0003-3305-6281]{Jeff Valenti}
\affiliation{Space Telescope Science Institute, 3700 Charles St., Baltimore, MD 21218}
\email{valenti@stsci.edu}

\author[orcid=0000-0003-2415-2191]{Julien de Wit}
\affiliation{Department of Earth, Atmospheric and Planetary Science, Massachusetts Institute of Technology, 77 Massachusetts Avenue, Cambridge, MA 02139, USA}
\email{jdewit@mit.edu}

\author[orcid=0009-0009-3020-3435]{Valeriy Vasilyev}
\affiliation{Max Planck Institute for Solar System Research, Justus-von-Liebig-Weg 3, 37077 G¨ottingen, Germany}
\email{vasilyev@mps.mpg.de}

\author[orcid=0000-0001-5989-7594]{Chia-Lung Lin}
\affiliation{Steward Observatory, The University of Arizona, 933 N. Cherry Avenue, Tucson, AZ 85721, USA}
\email{chialunglin@arizona.edu}


\author[orcid=0000-0003-3714-5855]{D\'aniel Apai}
\affiliation{Steward Observatory, The University of Arizona, 933 N. Cherry Avenue, Tucson, AZ 85721, USA}
\affiliation{Lunar and Planetary Laboratory, University of Arizona, 1629 E. University Boulevard, Tucson, AZ 85721, USA}
\affiliation{Alien Earths Team, NASA ICAR/NExSS, USA}
\email{apai@arizona.edu}

\author[orcid=0000-0002-5322-2315]{Ana Glidden}
\affiliation{Department of Earth, Atmospheric and Planetary Science, Massachusetts Institute of Technology, 77 Massachusetts Avenue, Cambridge, MA 02139, USA}
\affiliation{Kavli Institute for Astrophysics and Space Research, Massachusetts Institute of Technology, Cambridge, MA 02139, USA}
\email{aglidden@mit.edu}

\author[orcid=0000-0002-8052-3893]{Prajwal Niraula}
\affiliation{Department of Earth, Atmospheric and Planetary Science, Massachusetts Institute of Technology, 77 Massachusetts Avenue, Cambridge, MA 02139, USA}
\affiliation{Kavli Institute for Astrophysics and Space Research, Massachusetts Institute of Technology, Cambridge, MA 02139, USA}
\email{pniraula@mit.edu}

\author[orcid=0000-0002-8864-1667]{Peter Plavchan}
\affiliation{George Mason University, 4400 University Drive, Fairfax, VA 22030, USA}
\email{pplavcha@gmu.edu}

\author[orcid=0000-0002-3627-1676]{Benjamin V. Rackham}
\affiliation{Department of Earth, Atmospheric and Planetary Science, Massachusetts Institute of Technology, 77 Massachusetts Avenue, Cambridge, MA 02139, USA}
\affiliation{Kavli Institute for Astrophysics and Space Research, Massachusetts Institute of Technology, Cambridge, MA 02139, USA}
\email[]{brackham@mit.edu}

\author[orcid=0000-0003-3989-5545]{Noah Tuchow}
\affiliation{Steward Observatory, The University of Arizona, 933 N. Cherry Avenue, Tucson, AZ 85721, USA}
\email{nwtuchow@arizona.edu}

\author[orcid=0000-0003-0156-4564]{Luis Welbanks}
\affiliation{School of Earth and Space Exploration, Arizona State University, Tempe, AZ, USA}
\email[]{luis.welbanks@asu.edu}

\begin{abstract}

Stellar photospheric heterogeneities, such as starspots and faculae, are a fundamental limitation for exoplanet transmission spectroscopy. Inhomogeneous surfaces can imprint wavelength-dependent signals during transits that may mimic or mask atmospheric absorption features, especially for planets orbiting cool low-mass stars. Recent work has shown that the information content of transmission spectroscopy observations can be sufficient to correct for stellar contamination, but only if stellar photosphere and active-region models have adequate fidelity. This requires empirical benchmarking with observations that validate next-generation stellar models and identify which spectral diagnostics best encode heterogeneity properties as a function of spectral type, activity level, and time. We propose a two-generation Nautilus program that leverages the scalable architecture of the observatory concept. Generation 1 would use broad-wavelength time-series observations of transiting exoplanet systems to connect starspot-crossing events and out-of-transit variability to localized and disk-integrated heterogeneity properties. Generation 2 would use the optimized spectral diagnostics identified in Generation 1 to conduct slitless spectroscopic monitoring of large samples of GKM stars on different timescales. Generation 2 instrumentation would include activity tracers of both the photosphere and chromosphere. This program would provide the benchmark data and population-level framework needed to turn stellar contamination into a calibrated input for exoplanet atmospheric retrievals.

\end{abstract}


\section{Opening Statement}

This White Paper presents a potential science case for the Nautilus Space Observatory, a concept under development for a NASA Strategic Mission for the Astro 2030 Decadal Survey. Nautilus is a constellation of space telescopes and will provide a modular, scalable, sustainable, upgradable, expandable space observatory that can be deployed rapidly and then expanded progressively. The core concept for Nautilus is described in \cite{apai19}. This White Paper is part of the first series of science white papers capturing ideas that emerged from the Nautilus Science Case workshop (held at MIT in May 2026).

\section{Scientific Context and Problem Statement} 

The characterization of planetary atmospheres ranging from rocky worlds to ultra-hot Jupiters provides us with the information needed to address fundamental questions of planet climate. Despite the improved sensitivity and precision of JWST, recent observations of planets around low-mass M dwarfs and active GKM stars have featureless transmission spectra. The interpretation of these spectra are often viewed through three potential lenses: (I) the planet lacks a substantial atmosphere; (II) the planet has a high mean-molecular weight atmosphere; or (III) the planet's atmosphere is masked or mimicked by the transit light source effect \citep[TLSE;][]{rackham18, rackham19} --- an effect where unocculted stellar heterogeneities (e.g. starspots, faculae, plages) contaminate an observed transmission spectrum. Importantly, stellar contamination is not necessarily an irreducible noise floor. Recent work suggests that transmission spectroscopy observations contain enough information to correct for stellar contamination \textit{if} the stellar photosphere and active-region models have sufficient fidelity \citep[e.g.][]{Rackham24}. Time-resolved spectroscopy offers an empirical path towards benchmarking the next generation of photosphere and heterogeneity spectra directly \citep{Berardo2024}. These studies highlight the importance of observing active stars to obtain the benchmark data needed to validate stellar models and identify the absorption lines and/or bands that are the most sensitive to the properties of photospheric heterogeneities.

Measuring the properties of unocculted heterogeneities is a field-wide challenge that needs to be addressed for both exoplanet and stellar astrophysics. The presence of stellar heterogeneities has been informed through long-baseline photometric monitoring \citep[e.g.,][]{berdyugina05, morris20}, evidence of starspot crossing events via exoplanet transits \citep[e.g.,][]{sing11, fu22, ahrer25}, and spectroscopic observations that can only be explained by the presence of multiple temperature components \citep[e.g.,][]{neff95, gully17}. While stellar heterogeneities have been observed, there are still many open questions about the radii, longitudinal, and latitudinal distribution of heterogeneities, the temperatures of different photospheric features, and the timescale on which photospheric features appear, change, or disappear. Understanding these fundamental properties requires a program that first calibrates stellar heterogeneity diagnostics using information-rich, broad-wavelength, time-resolved observations of benchmark transiting systems and subsequently extends those diagnostics to large stellar samples. In this framework, a first-generation Nautilus program would target known active stars with transiting exoplanets to identify the wavelength regions where the information content of photospheric heterogeneities is densest across spectral type, while a second-generation program would use informed photospheric and chromospheric activity tracers to measure and monitor heterogeneity properties across a statistically large sample of stars across the HR diagram.



\section{Science Objectives} 


The primary objective of this program is to obtain the benchmark stellar data needed to move exoplanet transmission spectroscopy beyond the model-limited regime imposed by stellar contamination. The central challenge to overcome this roadblock is to validate and calibrate current photosphere and heterogeneity models with observations that directly connect localized stellar heterogeneities to disk-integrated spectral diagnostics. Using a combination of transit and non-transit observations, the Nautilus Space Observatory would address this challenge by answering the following questions:

\begin{enumerate}

\item \textbf{What physics are missing in current photospheric models, and how do observations of heterogeneous stellar surfaces constrain them?} Current stellar atmosphere models struggle to reproduce complex, non-local thermodynamic equilibrium effects. This failure mode is most acute for cool low-mass stars. Broad-wavelength, high signal-to-noise time-series observations of benchmark transiting systems will provide empirical tests of stellar models across photospheric and chromospheric diagnostics, identifying where current models fail as a function of spectral type and activity level. Benchmarking simulations with observations is required is we are to break through the ``stellar noise'' barrier in exoplanet transmission spectroscopy.


\item \textbf{Which spectral features encode the most information on heterogeneity properties across spectral types and activity levels?} Measuring the correlation between a star's fundamental parameters (e.g., mass, age, magnetic field strength, metallicity) and its surface features is critical for contextualizing planetary environments. However, not all wavelength regions contain equal information. A key deliverable of the first-generation Nautilus program will be to determine which spectral regions and activity indicators most strongly encode heterogeneity properties such as temperature and coverage fraction across the range of starspot, faculae, and granulation seen in GKM stars. This information-content map will define the optimal wavelength bands for the second generation slitless Nautilus survey.



\item \textbf{How do heterogeneities along the transit chord differ from disk-integrated features, and how do we reconcile these differences?} Exoplanet transits offer a unique magnifying glass to understanding spot and faculae properties along the transit chord through starspot crossing events (SCEs). Such observations can provide precise measurements of umbral and penumbral temperatures -- information that is often lost when only fitting stellar SEDs. However, the derived occulted properties may deviate from the disk-integrated properties. Pairing transit observations with out-of-transit monitoring establishes the empirical translation between occulted and unocculted heterogeneities, enabling stellar-contamination corrections to be applied beyond the small number of systems with transiting planets.

\item \textbf{How do photospheric heterogeneities evolve across timescales ranging from rotation to activity cycles, and how is this evolution imprinted on transmission spectra?} Stellar magnetic activity operates across a vast range of timescales: active regions emerge and decay over hours to days, rotation changes the observed disk-integrated surface over days to weeks, and activity cycles drive long-term evolution over years to decades. Each time the surface changes, it imprints a different signal into exoplanet transmission spectra. Time-series monitoring of benchmark transiting systems will quantify how active region properties change across these timescales, providing the empirical foundation needed to model and remove time-dependent stellar contamination.

\item \textbf{How do we scale calibrated heterogeneity diagnostics to population-level studies?} SCEs provide spatially resolved snapshots of a star's active surface, but this technique is limited to systems with transiting planets. Once the most informative photospheric and chromospheric tracers are identified, a second-generation Nautilus program can use slitless spectroscopic monitoring to measure stellar heterogeneity properties across a large sample of GKM stars. By anchoring localized SCE-derived properties to disk-integrated spectra, we can derive scaling laws applicable to stars without transiting planets, producing the population-level priors required for robust exoplanet atmospheric retrievals and for understanding how stellar activity varies across the HR diagram.


\end{enumerate}


\section{Data Requirements}


To advance our understanding of stellar photospheric heterogeneity and its impact on exoplanet transmission spectroscopy, we require a coordinated two-generation instrumentation strategy. The first generation provides the broad-wavelength, time-resolved benchmark data needed to validate next-generation stellar photosphere and active region models. It is also required to map which absorption lines and/or bands have the highest information content on heterogeneity properties as a function of spectral type. Following the success of the Generation 1, the second generation will be optimized to monitor a large sample of stars efficiently through slitless spectroscopy. The wavelength range of the Generation 2 spectrograph will be set by the results of Generation 1. This staged approach directly leverages the inherent scalability and multi-generational architecture of the Nautilus Space Observatory.

\subsection{Generation 1: Leveraging Systems with Transiting Exoplanets}

The initial phase of the program will focus on building a high-fidelity benchmark spectral library of a targeted sample of active stars with transiting exoplanets. These observations are required to isolate individual surface components such as spots, faculae, plages, and granulation, and to validate the stellar photosphere and active-region models needed for reliable stellar-contamination corrections. This stage requires a medium resolution ($R\sim1000$) single-slit spectrograph spanning a wide wavelength range of $\lambda=0.3-1.2\mu$m to simultaneously observe photospheric diagnostics and chromospheric activity tracers. Relevant features include absorption lines and molecular bands sensitive to cool photospheric components \citep[e.g., Fe~\textrm{I}, OH, SiO, MgH, TiO, H$_2$O;][]{polyansky97} as well as activity diagnostics (e.g., Balmer break, Ca~\textrm{II}~H~\&~K, H$\alpha$, Ca~\textrm{II}~infrared triplet). By deriving flux-calibrated high-resolution time-series optical spectra, we will measure how these diagnostics vary with the changing photospheric and chromospheric state of the star.


For Generation 1, we require observations that have rapid sampling to resolve the fine structure of SCEs. This requires exposure times on the order of $\leq 2$~minutes. The out-of-transit observations should continuously monitor targets with sub-pixel precision over a minimum of 6 stellar rotation periods. This is required to resolve rotational modulation and active-region evolution, which occurs on the timescale of $\sim3-4$~$P_\mathrm{rot}$ \citep{}. Rotation periods of the current sample of planet-hosting stars are $P_\mathrm{rot} =1.2-15$~days \citep[e.g.][]{rizzuto20, dong22, mallorquin23, thao24, vach25}. We require a precision of 10 ppm min$^{-1}$ in the derived white-light curve to accurately detect SCEs \citep{Murphy26}. 

We will map the thermal and spatial properties of heterogeneities through SCEs. Once completed, we will directly correlate the derived properties to measurable variations in specific absorption lines and/or molecular bands. Identifying how features like TiO or H$_2$O line profiles change under different SCE properties will enable the creation of an empirical framework of photospheric activity that can be applied to other stars. At present, it is unclear what spectral diagnostics provide the most robust mapping between localized heterogeneity properties and disk-integrated stellar spectra, and how this mapping varies across stellar parameters. This is why we require Generation 1 to have a wide wavelength range: it is designed to measure the wavelength-dependent time-variable properties of photospheric and chromospheric heterogeneities. Once the highest-information spectral regions are identified and validated against stellar models, the program can proceed to Generation 2.


\subsection{Generation 2: Statistically Measuring Heterogeneity Properties across Hundreds of Stars}


The first generation program will benchmark narrow wavelength ranges that are the most sensitive to photospheric heterogeneity properties and chromospheric activity. Once these diagnostics are calibrated and validated, Generation 2 could target these optimized bands across a statistical sample of stars to derive a population-scale census of heterogeneity properties. For Generation 2, we require a slitless spectrograph whose wavelength coverage is defined by the empirical calibrations completed in Generation 1. The slitless architecture enables efficient, simultaneous observations of many stars across a broad range of spectral types.
The trade-off for Generation 2 is the wavelength coverage must be relatively narrow ($<0.5\mu$m wavelength coverage) to minimize overlapping spectral traces on the detector. We require a lower resolution ($R\sim 100-500$) and longer cadence ($\sim 10-30$~minutes) than Generation 1, with similar precision ($50-100$\,ppm per spectral bin). Generation 2 has the same monitoring baseline and stability as Generation 1.

\section{Analysis and Interpretation} 



The analysis of the multi-generational Nautilus data will be designed to validate next-generation stellar photosphere and active region models, identify key spectral diagnostics that most strongly constrain stellar heterogeneity properties, and propagate those constraints into exoplanet atmospheric retrievals. In this framework, Generation 1 provides the benchmark data set for model validation and diagnostic calibration, while Generation 2 applies the calibrated diagnostics to population-scale stellar monitoring. For Generation 1, we observe a handful of benchmark transiting systems with known previous SCEs. While heterogeneities will likely change, we will target both known active stars and M dwarfs, which all but guarantee SCEs will be present. Starspot contrast spectra will be derived by measuring the change in amplitude of the SCE as a function of wavelength \citep[e.g.,][]{murray26}. We will fit the contrast spectra assuming the spots are comprised of umbral and penumbral regions with different covering fractions. This has been demonstrated for JWST \citep{Murphy26}. This allows us to accurately derive the temperature components of heterogeneities along the transit chord. Simultaneously, we will measure the changing line profile properties of specific absorption lines and molecular bands the out-of-transit stellar spectra. These empirical spectral indicators will be cross-correlated with the derived temperature components from the SCE modeling. By mapping these correlations, we will establish an empirical library of photospheric heterogeneity properties. Further, we will use these observations to advance 3D magnetohydrodynamic models of stellar photospheres and generate a library of new spectral models to compare future observations to.

For Generation 2, we will shift the analysis to a population-scale framework. Using the empirical calibrations established in Generation 1, we will monitor the variations of hundreds of targets within a narrow wavelength window. By correlating observed variations with precise photospheric and chromospheric diagnostic templates calibrated in Generation 1, we will self-consistently derive stellar heterogeneity components. This allows us to isolate the true spectrum of hundreds of stars across various spectral types, moving from individual stellar monitoring to population-scale statistics. 

A successful analysis is defined by three primary milestones. First, Generation 1 will deliver a benchmark validation data set for stellar photosphere and active region models, identifying where current models succeed or fail as a function of wavelength, spectral type, and activity level. Second, the program will identify a minimal set of high-information photospheric and chromospheric diagnostics that connect localized heterogeneity properties from SCEs to disk-integrated stellar spectra. Third, the validated models and empirical diagnostics will be propagated into exoplanet atmospheric retrievals to demonstrate a measurable reduction in TLSE-driven uncertainties. Success will be achieved when stellar contamination corrections are no longer limited by unvalidated stellar models, but by the intrinsic limitations of the data itself.

\section{Relevant Science Requirements}

We summarize the relevant science requirements for Generation 1 in Table~\ref{tab:gen1} and for Generation 2 in Table~\ref{tab:gen2}.

\section{Relevance to Nautilus and Mission Class} 

This science case is particularly well matched to the scalable, multi-generation architecture of the Nautilus Space Observatory. The central objective is to move exoplanet transmission spectroscopy beyond the model-limited regime imposed by stellar contamination. Achieving this requires an iterative program. First, we require broad-wavelength benchmark observations to validate stellar photosphere and active region models and identify the spectral regions with the highest heterogeneity information content. Second, we require population-scale monitoring in optimized diagnostic bands to extend these empirical calibrations across different stellar parameters.

The first generation of Nautilus observations would require high-stability, broad-wavelength, time-series spectroscopy of selected transiting systems. These data would provide the benchmark measurements needed to connect localized starspot-crossing events, disk-integrated variability, and stellar model predictions. This phase benefits from large collecting area, stable space-based spectroscopy, and wavelength coverage that is difficult to obtain simultaneously with existing facilities. Its purpose is to determine which photospheric and chromospheric diagnostics are most informative for stellar-contamination corrections.

The second generation would then use the results of the first generation to define a narrower, optimized bandpass for slitless spectroscopic monitoring. This stage is where the constellation architecture becomes especially powerful. Multiple Nautilus units could monitor many fields in parallel over baselines spanning several stellar rotation periods ($\sim$\,months), enabling a statistical census of stellar heterogeneity and activity states across GKM stars in a relatively short period of time. The ability to fly later units with instrument configurations informed by earlier benchmark observations is a distinctive advantage of Nautilus: the observatory can learn which diagnostics matter most before scaling to the population level.

A constellation is therefore not merely an efficiency gain for this science case: it is part of the science strategy. Parallel observing capability increases the number of stars that can be monitored, while the multi-generation architecture allows the mission to transition from model validation to optimized survey execution. This staged approach would provide the empirical foundation needed by Nautilus \citep{welbanks_wp, pascucci_wp}, JWST, Habitable Worlds Observatory \citep{}, ARIEL \citep{}, and other future exoplanet transmission spectroscopy facilities.

This program could begin at a Probe-class scale through a focused Generation 1 benchmark survey of selected transiting systems. However, the full science return---a population-level calibration of stellar heterogeneity diagnostics across stellar type, activity level, and time---requires the expanded observing efficiency of a Flaglet- or Flagship-class Nautilus constellation. In this sense, the stellar-heterogeneity program provides a natural maturation path: a smaller initial implementation can validate the methodology and requirements, while the larger constellation enables the statistical survey needed to support robust exoplanet atmospheric demographics.





\section{Relevance to NASA and Astrophysics Strategy} 

This science case is directly aligned with NASA's strategic priority of understanding worlds beyond the Solar System and assessing their potential habitability. The Astro2020 Decadal Survey identifies the search for habitable worlds and life beyond the Solar System as a central theme for the coming decades, motivating both the Habitable Worlds Observatory and the broader development of facilities capable of robust exoplanet atmospheric characterization. A key prerequisite for that goal is understanding the stars that host these planets.

We are in an era where our interpretation of exoplanet transmission spectra are model-limited, rather than photon-limited. As JWST and future missions push toward smaller planets, we need to measure the stellar photospheric heterogeneities that can mimic, mask, or distort the atmospheric signals. This is seemingly ancillary science is central to NASA's exoplanet science goals. Empirically calibrated and validated stellar photosphere and active region models are therefore enabling infrastructure for the search for habitable worlds.

The Nautilus Space Observatory would address this strategic need by providing the benchmark observations required to validate next-generation stellar models and identify the photospheric and chromospheric diagnostics most relevant to exoplanet atmospheric retrievals. This program directly supports the interpretation of JWST transmission spectra, informs the observing and analysis strategies of complimentary observations and future missions, and provides essential context for key targets for the Habitable Worlds Observatory.

This science case is also well aligned with the goals of NASA's ASTRA Initiative, which seeks to mature strategic mission concepts before the next Decadal Survey. The proposed two-generation Nautilus program links a clear astrophysical bottleneck to a scalable mission architecture. First, we will benchmark and validate the stellar models and diagnostics required for robust atmospheric inference. Next, we will extend those diagnostics to population-scale stellar monitoring. In doing so, Nautilus would ensure that future exoplanet surveys are limited by observational precision and an understanding atmospheric physics, rather than by unvalidated stellar models.

\begin{deluxetable}{lll}
\tabletypesize{\scriptsize}
\tablecaption{Generation 1: Single-slit wide-wavelength spectrograph to study the stellar heterogeneities on a smaller sample of stars.\label{tab:gen1}}
\tablewidth{0pt}
\tablehead{
\colhead{Requirement} & \colhead{Range} & \colhead{Science Driver}
}
\startdata
Photometric Filters & N/A & \\
Wavelength Coverage [$\mu$m] & $0.3-1.2$ & \\
Target Brightness [mag] & V$\leq18$ & \parbox[t]{4in}{We need to understand heterogeneity properties on a range of stars, from fully convective M dwarfs to solar analogs. To accommodate for this range of stellar properties, we require the ability to obtain high precision spectroscopic observations of stars ranging down to $V \leq 18$.} \\
Min. Photom. Precision [ppm] & 10 min$^{-1}$ & \parbox[t]{4in}{We require the precision to resolve potential starspot crossing events and the substructure of those events. As such, we require high-precision white-light curves on the order of 10~ppm~min$^{-1}$ when the entire spectrum is summed.}  \\
Image Res. [diff. limit] & N/A & \\
Min. Sky Coverage [deg$^2$] & $<0.1^\circ\times0.1^\circ$ & \parbox[t]{4in}{We require observations of single targets without contamination from nearby targets. To this end, we require a very small field of view.}  \\
Min. Contrast & N/A &  \\
Spectral Resolving Power & $1000$ at $0.4\mu$m & \parbox[t]{4in}{We require the ability to measure the slope of the stellar spectral energy distribution in addition to resolving individual absorption and/or emission features (e.g., Ca~\textrm{II}~H~\&~K, H~$\alpha$).} \\
\hline
Relevant Timescales [s] & $\sim10^5$ & \parbox[t]{4in}{} \\
Monitoring Baseline [d] & $\sim$~months & \parbox[t]{4in}{We will monitor stars for $\sim$6 rotation periods, which can range from weeks to months.} \\
Cadence [s] & $120 \pm 60$ &  \parbox[t]{4in}{We require a cadence that is sufficiently fast to resolve starspot crossing events. SCEs typically last on the order of $> 5$~minutes. Thus, we need a cadence of $\sim120$~s to resolve these events.} \\
Rapid Response Time [s] & N/A & \parbox[t]{4in}{} \\
\hline
Data Volume &   & \parbox[t]{4in}{} \\
Pointing Precision [arcsec] & $<10''$ & \parbox[t]{4in}{} \\
\enddata
\end{deluxetable}

\begin{deluxetable}{lll}
\tabletypesize{\scriptsize}
\tablecaption{Generation 2: Slitless spectroscopy over a wide field-of-view to enable statistical constraints on stellar heterogeneities.\label{tab:gen2}}
\tablewidth{0pt}
\tablehead{
\colhead{Requirement} & \colhead{Range} & \colhead{Science Driver}
}
\startdata
Photometric Filters & N/A & \\
Wavelength Coverage [$\mu$m] & TBD & \parbox[t]{4in}{The precise wavelength range will be determined based on the results of Generation 1.} \\
Target Brightness [mag] & V$\leq18$ & \parbox[t]{4in}{Same as Generation 1.} \\
Min. Photom. Precision [ppm] & 10 min$^{-1}$ & \parbox[t]{4in}{Same as Generation 1.}  \\
Image Res. [diff. limit] & N/A & \\
Min. Sky Coverage [deg$^2$] & $<0.6^\circ\times0.6^\circ$ & \parbox[t]{4in}{We require observations of multiple stars simultaneously. To this end, we require a larger field-of-view compared to Generation 1.}  \\
Min. Contrast & N/A &  \\
Spectral Resolving Power & $500$ at $\lambda=$TBD & \parbox[t]{4in}{In Generation 2, we will focus on a smaller wavelength range to enable slitless spectroscopy of dozens of stars simultaneously.} \\
\hline
Relevant Timescales [s] & $\sim10^5$ & \parbox[t]{4in}{} \\
Monitoring Baseline [d] & $\sim$~months & \parbox[t]{4in}{Same as Generation 1} \\
Cadence [s] & $1200 \pm 600$ &  \parbox[t]{4in}{We require a cadence to measure variations in line profiles across a stellar rotation period. Rotation periods can range from $\geq 0.5$~days. Thus, we need sub-hour cadence to resolve changes in the heterogeneities.} \\
Rapid Response Time [s] & N/A & \parbox[t]{4in}{} \\
\hline
Data Volume & $\sim$TBs  & \parbox[t]{4in}{We require the raw images from Generation 2 to accurately extract the spectra of $\sim10$s of stars in any given field of view. The locations of the spectra on the detector will change depending on the field of view. Thus, it is most reliable to work with the raw images themselves, rather than relying on onboard processing.} \\
Pointing Precision [arcsec] & $<10''$ & \parbox[t]{4in}{} \\
\enddata
\end{deluxetable}


\begin{acknowledgments}
We thank the Heising-Simons Foundation for supporting the Nautilus Science Case Workshop. 
\end{acknowledgments}

\begin{contribution}

A.~D.~Feinstein and J.~Valenti conceptualized the first generation Nautilus science case presented in this work. J.~de~Wit conceptualized the second generation Nautilus science case presented in this work. A.~D.~Feinstein and J.~de~Wit wrote the bulk of the manuscript. V.~Vasilyev, C.-L.~Lin participated in helpful discussions that lead the to development of this science case.  D.~Apai, A.~Glidden, P.~Niraula, P.~Plavchan, B.~V.~Rackham, N.~Tuchow and L.~Welbanks reviewed and provided comments on the manuscript. 

\end{contribution}



\bibliography{sample701}{}
\bibliographystyle{aasjournalv7}

\end{document}